# Dynamic modulation of photonic crystal nanocavities using gigahertz acoustic phonons

Daniel A. Fuhrmann[1,2], Susanna M. Thon[3], Hyochul Kim[3,4], Dirk Bouwmeester[3,5], Pierre M. Petroff[2], Achim Wixforth[1,6] & Hubert J. Krenner[1,*]

(1) Lehrstuhl für Experimentalphysik 1 and Augsburg Centre for Innovative Technologies (ACIT), Universität Augsburg, Universitätsstr. 1, 86159 Augsburg, Germany

(2) Materials Department, University of California, Santa Barbara CA 93106, United States

(3) Physics Department, University of California, Santa Barbara CA 93106, United States

(4) Department of Electrical and Computer Engineering, IREAP, University of Maryland, College Park MD 20742, United States

(5) Huygens Laboratory, Leiden University, P.O. Box 9504, 2300 RA Leiden, The Netherlands

(6) Center for Nanoscience (CeNS), Ludwig-Maximilians-Universität München, Geschwister-Scholl-Platz 1, 80539 München, Germany

\*       hubert.krenner@physik.uni-augsburg.de

**Photonic crystal membranes (PCM) provide a versatile planar platform for on-chip implementations of photonic quantum circuits[1-3]. One prominent quantum element is a coupled system consisting of a nanocavity and a single quantum dot (QD)[4-7] which forms a fundamental building block for elaborate quantum information networks[8-10] and a cavity quantum electrodynamic (cQED) system controlled by single photons[3]. So far no fast tuning mechanism is available to achieve control within the system coherence time. Here we demonstrate dynamic tuning by monochromatic coherent acoustic phonons formed by a surface acoustic wave (SAW) with frequencies exceeding 1.7 gigahertz, one order of magnitude faster than alternative approaches[5-7]. We resolve a periodic modulation of the optical mode exceeding eight times its linewidth, preserving both the spatial mode profile and a high quality factor. Since PCMs confine photonic and phononic excitations[11,12], coupling optical to acoustic frequencies, our technique opens ways towards coherent acoustic control of optomechanical crystals.**

In basic research SAWs found applications in the investigation of fundamental quantum effects in nanosystems[13-17], the manipulation of photonic bandgap structures[18], microcavity and surface plasmon polaritons[19-21] with frequencies spanning from a few megahertz up to a several gigahertz. We electrically generate SAWs by applying a short radio frequency (RF) voltage pulse to interdigital transducer electrodes (IDT) as shown schematically in Fig. 1 (a). As this pulse propagates across the PCM, it dynamically deforms the region of an optimized, high quality defect cavity[22] on timescales defined by the period of the SAW ($T_{SAW}$). Thus, the nanocavity is dynamically stretched at $t = $ ¼ $T_{SAW}$ and compressed half a period later ¾ $T_{SAW}$ on a nanometre scale which is accompanied by a dynamic spectral tuning of the nanocavity mode. For the expanded situation, the effective length of the cavity is increased and we expect the cavity mode to be red-shifted to longer wavelengths at $t = $ ¼ $T_{SAW}$. In contrast, for the compressed cavity at $t = $ ¾ $T_{SAW}$ the situation is reversed, the effective length of the cavity is reduced and, accordingly, the spectral resonance should be blue-shifted. To experimentally confirm this tuning, we fabricated hybrid PCM-SAW devices directly on GaAs in a two-step process and monitor the nanocavity mode via the optical emission of an ensemble of self-assembled QDs. In Fig. 1 (b), we present a typical time-integrated emission spectrum in the range of the QD emission which is decorated by a Lorentzian cavity mode with a high quality ($Q$-) factor of $Q_0 \sim 8300$ at $\lambda_c = 911.1$ nm. When a SAW is generated at $f_{SAW} = 1.703$ GHz by applying a RF power of $P_{RF} = +17$ dBm to the IDT, the initially sharp cavity resonance (solid black line) significantly broadens in the time-integrated spectrum to a full width half maximum, $\Delta\lambda_c = 0.57\pm0.1$ nm, and its amplitude decreases (red line). Due to the non-Lorentzian lineshape direct fitting is not possible and the uncertainty of $\Delta\lambda_c$ becomes limited by the full width half maximum (FWHM) of the unperturbed cavity. While the decrease of the integrated intensity results from less efficient carrier injection into and reactivation out of the QDs[17], the broadening arises from the dynamic spectral tuning of the cavity mode. Since this data is averaged over $> 10^{10}$ SAW cycles the observed FWHM provides a direct measure for the modulation range.

To confirm that the observed broadening indeed arises from mechanical actuation of the PCM by the SAW, we set $P_{RF} = +28$ dBm and scan the RF frequency through the design frequency of the IDT, $f_{SAW} = c_{SAW}/p$, where $p$ is the periodicity. We measure the FWHM of the cavity emission as a function of the applied frequency and plot the obtained values with their corresponding extraction uncertainty as symbols in Fig. 1 (c). Clearly, a broadening of the cavity emission is only detected in the narrow frequency band $f_{SAW} = 1703 \pm 2 \, \text{MHz}$. This remarkably narrow band corresponds to <0.2% of the IDT's resonance frequency. Furthermore, we independently confirm that this resonance indeed corresponds to SAW generation by the IDT. For the IDT converting the applied RF power into a SAW, we expect a reduction of the reflected RF intensity from the electrical circuit connecting to the IDT. This is nicely observed as a dip in the reflected signal (S11) in the experimental data plotted as a line in Fig. 1 (c) which overlaps almost perfectly with the measured broadening of the cavity emission. We performed similar experiments for three different SAW frequencies $f_{SAW}$ = 414 MHz, 850 MHz and 1.7 GHz and find the same narrow band response of the cavity mode exactly overlapping with the SAW excitation (see Supplementary Information). This unambiguously proves that the electrically generated coherent acoustic phonon strain and stress fields of the SAW spectrally tune the nanocavity photonic mode. Furthermore, our observation confirms nicely acoustic tuning performed on one-dimensional Bragg microcavities using SAW[20] or optically generated, broadband picosecond strain pulses[23]. Since the latter are generated on the backside of the sample and propagate as bulk acoustic waves, these cannot be directly applied to PCM-based systems.

Armed with this tuning mechanism, we can estimate the amplitude of the SAW from the observed broadening by a simple model considering the nanocavity as a one-dimensional Fabry-Perot (*FP*-) resonator with oscillating separation between the two mirrors (Inset Fig. 1b). The resonance wavelength of such a *FP*-resonator is given by $\lambda_{FP} = \frac{2d}{j}$, with $d$ being the separation between the two mirrors and $j$ the index of the mode. Our *L3* cavity design[22] consists of 3 missing holes which translates to an effective cavity length of $2.5 \cdot \lambda$ and $j = 5$. Thus, the variation of resonance wavelength in vacuum is given $\Delta\lambda_c = \frac{n_{eff} \cdot \Delta d}{2.5}$, accounting for the effective refractive index of the membrane $n_{eff} \sim 2.75$. From these considerations we can estimate the SAW amplitude to $A_{SAW} = \frac{\Delta d}{2} = 0.32$ nm, in good agreement with realistic values.

We probe the emission of three different nanocavities modes and SAW frequencies as a function of the RF power which is proportional to the acoustic power $P_{SAW} \propto P_{RF}$ and plot the detected emission spectra as a function of $P_{RF}$ in false colour representation in Fig. 2 (a)-(c). At low RF powers we observe the unperturbed emission of the nanocavity mode which shows a pronounced and continuous broadening as $P_{RF}$ increases for all three frequencies up to a maximum of $\Delta\lambda_c > 1.5$ nm at the highest power levels which corresponds to $A_{SAW} = 0.55$ nm using the simple *FP*-resonator model. Since $A_{SAW} \propto \sqrt{P_{SAW}}$, the extracted FWHM of the cavity emission should exhibit a linear dependence when plotted as a function of $\sqrt{P_{RF}}$. Such an anticipated linear increase is indeed observed experimentally for all three frequencies studied as shown in Fig. 2 (d).

In order to describe our experimental findings, we extend the *FP*-model and apply three-dimensional finite difference time domain (3D-FDTD) simulations. We start by taking the optimized *L3* cavity design[22] of our devices and superimpose the lateral distortion of the PCM by the SAW. The effect of the vertical deformation can be neglected to first order since it only weakly affects the waveguided mode in the PCM. We also do not account for the weak $\sim 10^{-4}$ variation[24] of the refractive index of GaAs ($n \sim 3.4$) due to its large contrast to air. Then, we calculate the dynamic modulation of the cavity resonance $\lambda_c$ and *Q*-factor for a fixed SAW wavelength ($\lambda_{SAW}$) and amplitude ($A_{SAW}$) by treating the deformation as a quasi-static perturbation and apply the same phase conversion as for the *FP*-model [cf. Fig. 1 (a)]. In Fig. 3 (a), we present an example for the calculated periodic modulation of $\lambda_c$ (solid line) and *Q*-factor (red line) for $A_{SAW} = 1$ nm and $\lambda_{SAW} = 7a$, $a = 0.26$ μm being the lattice constant of the PCM. $\lambda_c$ nicely follows the sinusoidal modulation determined by the geometric deformation, which is anticipated by the *FP*-model. Moreover, we would expect that in the limit of our experiments $\left(\Delta d/d \ll 1\right)$ the *Q*-factor remains constant. This expectation is confirmed by the 3D-FDTD calculations, which predict a weak $\Delta Q/Q_0 \approx 3\%$ modulation in the order of the resolution of these simulations. We extend this 3D-FDTD approach and assess the dynamic tuning range, $\Delta\lambda_c$, as a function of $A_{SAW}$ and $\lambda_{SAW}$ presented in Fig. 3 (b) and (c), respectively. Remarkably, using this accurate method, the calculation of the amplitude dependence of the spectral shift clearly reproduces the *linear* tuning, which we anticipated within the simple *FP*-model and observed in the experimental data [cf. Fig. 2 (d)] for all values of $\lambda_{SAW}$. However, in contrast to a *FP*-resonator, the model predicts that the tuning range depends on $\lambda_{SAW}$, giving rise to the different slopes in Fig. 3 (b). We evaluate this effect in Fig. 3 (c) by plotting the calculated tuning range $\Delta\lambda_c$ for a fixed value of $A_{SAW} = 1.0$ nm as a function of $\lambda_{SAW}$. We find that the tuning range exhibits a maximum at $\lambda_{SAW} = 7a$, which decreases exponentially for both increasing and decreasing $\lambda_{SAW}$. This can be understood qualitatively as the length of the optimized *L3*-cavity of ~ $3.4a$ compares well to $\lambda_{SAW}/2$ giving rise to the maximum relative deformation of the nanocavity as a whole. Finally, we evaluate the spatial electro-magnetic field profile within the cavity for maximum extension ($t=0.25\ T_{SAW}$) and compression ($t=0.75\ T_{SAW}$). The two profiles are compared in Fig. 3 (d) on the left- and right-hand side, respectively. Clearly, no shifts of the spatial mode profile and in particular of the central anti-node are resolved in our calculations. Such a stationary mode profile is a crucial requirement for both dynamic QD-nanocavity coupling and coherent acoustic control of optomechanical crystals.

For the data presented up to this point, we recorded *time-integrated* spectra averaging over the entire SAW cycle. To resolve the high-frequency dynamic nature of our SAW based tuning mechanism, we now employ *phase-correlated stroboscopic spectroscopy*[25] by setting the laser repetition time to $\Delta t_{laser} = n \cdot T_{SAW}$, $n$ being an integer. In this type of experiment, we excite the system at a *fixed time during the SAW cycle* and then perform time-integrated detection of the emission. A typical example of such a temporal scan of the emission of a nanocavity mode is presented in Fig. 4 (a) for $f_{SAW} = 414$ MHz excited with $P_{RF} = +32$ dBm. Without a SAW present, we determined the nanocavity emission wavelength and unperturbed *Q*-factor to be $\lambda_{c,0} = 917.45$ nm and $Q_0 \approx 5 \cdot 10^3$, respectively. In the experimental data, the modulation of the cavity resonance with the fundamental period of the SAW ($T_{SAW}$) and a bandwidth of $\Delta\lambda_c = 1.2$ nm is clearly resolved. This is in good agreement with the broadening observed in phase-averaged experiments. In Fig. 4 (b), we plot the extracted spectral position of the cavity resonance (symbols) which we find to be in excellent agreement with FDTD simulations for $A_{SAW} = 1.9$ nm (line). From this data we extract the relative temporal variation of the *Q*-factor given by $\frac{Q(t) - Q_0}{Q_0}$ plotted in Fig. 4 (b). It shows, in strong contrast to the resonance wavelength, a modulation with $2\ T_{SAW}$. At $t = 0$ and $t = \frac{1}{2}\ T_{SAW}$ at which the deformation of the cavity region is minimum, we find the maximum reduction of $Q(t)$. At times $t = \frac{1}{4}\ T_{SAW}$ and $t = \frac{3}{4}\ T_{SAW}$ for which the cavity region is maximally deformed the extracted *Q*-factor is not affected by the SAW and identical to $Q_0$. These counterintuitive observations do not arise from a degradation of the *Q*-factor but from the employed stroboscopic technique: We investigate a dynamic spectral shift with a fast tuning rate $|d\lambda_c / dt|$ which gives rise to an additional broadening of the cavity line. This can be estimated as $\delta\lambda_c \approx |d\lambda_c / dt| \cdot \tau$ with $\tau$ being the overall temporal resolution. We calculate the expected modulation of the linewidth and the nominal *Q*-factor (see Supplementary Information) for the given $f_{SAW}$ for the base time resolution of our system ($\tau_{system} \sim \tau_{laser} \sim 90$ ps) determined by the laser pulse width. This sets the maximum resolution for the $Q$-factor evaluation. A direct comparison between this model (black dashed line) and the experimental data in Fig. 4 (b) shows that the observed modulation is indeed dominated by this instrumental limitation. Taking into account the Purcell-enhancement of the radiative decay time of the QD emission[5,26] ($\tau_{rad} \sim 50$ ps) we obtain for $\tau_{system} + \tau_{rad} = 140$ ps, we find excellent agreement between the modelled $Q$-factor modulation (solid line) and the experimental data. Moreover, for long radiative decays our model predicts a significantly increased contrast of the $Q$-factor modulation as illustrated for $\tau = 500$ ps (dotted line) which is still shorter than the typical decay time of 800 ps of self-assembled QDs in an isotropic optical medium[26]. To confirm this analysis, we extract the $Q$-factor modulation from the 3D-FDTD calculations, which nicely reproduce the observed spectral tuning. Clearly, the calculated variation of the $Q$-factor plotted on the same scale (red line) in Fig. 4 (b) is indeed weak and large $Q$-values are maintained over the entire SAW cycle. Thus, our experimental data provides direct evidence for a large tuning range corresponding to more than 5.7 cavity linewidths without significant degradation of its $Q$-factor.

The broadband, RF spectral tuning of our SAW/PCM system will result in deeper insights in cQED, due to its fast tuning rates compared to lifetime of the emitters and photons inside the cavitiy[27]. The accessible tuning range is further expanded by the additional modulation of the QD emission[16,17]. Variation of the SAW frequency offers the possibility to investigate geometric resonances which enhance the tuning range or dynamic couple localized photonic and phononic modes[11,12,28], which can be also achieved on non-piezoelectric substrates, in particular Silicon[29]. One particularly tantalizing approach is to use the coherent monochromatic acoustic phonon field of a SAW for the coherent control of optomechanical crystals which in contrast to planar Bragg microcavities can be realized with sufficiently high quality on a PCM platform[11,12,30,31].

**Methods:**

*Heterostructure*

We start by fabricating the photonic crystal membranes from a semiconductor heterostructure grown by molecular beam epitaxy. This heterostructure consists of a 134 nm GaAs layer with self-assembled InAs QDs in its centre on top of a 920 nm thick $Al_{0.7}Ga_{0.3}As$ sacrificial layer. First, the photonic crystal structure is defined by electron beam lithography and transferred into the heterostructure by dry chemical ICP-RIE etching. Next, we remove the sacrificial layer wet chemically using hydrofluoric acid and release a fully suspended membrane. In the following step Titanium/Aluminium IDTs with design frequencies of $f_{SAW}$ = 414 MHz, 850 MHz and 1.7 GHz are defined by electron beam lithography and finalized in a conventional lift-off process at a distance of 1 mm from the PCM. The corresponding SAW wavelengths are $\lambda_{SAW} = 28a$, $14a$ and $7a$, with $a = 0.26$ μm being the PCM lattice constant.

*Experimental setup*

In our photoluminescence setup QDs in the PCM are excited by an externally triggered diode laser emitting ~90 ps pulses at a wavelength of 850 nm which is focused to a 2 μm spot using a 50x microscope objective. The emission is dispersed by a 0.5 m imaging grating monochromator and detected by a liquid $N_2$-cooled charge coupled device. The sample is cooled to $T$ = 5 K in a Helium-flow cryostat with integrated RF connections. The triggering scheme for phase-correlated stroboscopic spectroscopy used to resolve the full dynamics is described in detail in reference 25.

References:

1 Notomi, M., Shinya, A., Mitsugi, S., Kuramochi, E. & Ryu, H-Y. Waveguides, resonators and their coupled elements in photonic crystal slabs. *Optics Express* **12**, 1551-1561 (2004).

2 O'Brien, J. L., Furusawa, A. & Vuckovic, J. Photonic quantum technologies. *Nature Photon.* **3**, 687-695 (2009).

3 Faraon, A., Fushman, I., Englund, D., Stoltz, N., Petroff, P. & Vuckovic, J. Coherent generation of non-classical light on a chip via photon-induced tunnelling and blockade. *Nature Phys.* **4**, 859-863 (2008).

4 Noda, S., Fujita, M. & Asano, T. Spontaneous-emission control by photonic crystals and nanocavities. *Nature Photon.* **1**, 449-458 (2007).

5 Hennessy, K., Badolato, A., Winger, M. Gerace, D., Atatüre, M., Gulde, S., Fält, S., Hu, E. L. & Imamoglu, A. Quantum nature of a strongly coupled single quantum dot-cavity system. *Nature* **445**, 896-899 (2007).

6 Faraon, A., Majumdar, A., Kim H., Petroff, P. & Vučković, J. Fast Electrical Control of a Quantum Dot Strongly Coupled to a Photonic-Crystal Cavity. *Phys. Rev. Lett.* **104**, 047402 (2010).

7 Nomura, M., Kumagai, N., Iwamoto, S., Ota, Y. & Arakawa, Y. Laser oscillation in a strongly coupled single-quantum-dot-nanocavity system. *Nature Phys.* **6**, 279-283 (2010).

8 Imamoglu, A., Awschalom, D. D., Burkard, G., DiVincenzo, D. P., Loss, D., Sherwin, M. & Small, A. Quantum Information Processing Using Quantum Dot Spins and Cavity QED. *Phys. Rev. Lett.* **83**, 4204-4207 (1999).

9 Kok, P., Munro, W. J., Nemoto, K., Ralph, T. C., Dowling, J. P. & Milburn, G. J. Linear optical quantum computing with photonic qubits. *Rev. Mod. Phys.* **79**, 135-174 (2007).

10 Cirac, J. I., Zoller, P., Kimble, H. J. & Mabuchi H. Quantum State Transfer and Entanglement Distribution among Distant Nodes in a Quantum Network. *Phys. Rev. Lett.* **78**, 3221-3224 (1997).

11 Eichenfield, M., Chan, J., Camacho, R. M., Vahala, K. J. & Painter, O. Optomechanical crystals. *Nature* **462**, 78-82 (2009).

12 Mohammadi, S., Eftekhar, A. A., Khelif, A. & Adibi, A. Simultaneous two-dimensional phononic and photonic band gaps in opto-mechanical crystal slabs. *Optics Express* **18**, 9164-9172 (2010).

13 Wixforth, A., Kotthaus, J. P. & Weimann, G. Quantum Oscillations in the Surface-Acoustic-Wave Attenuation Caused by a Two-dimensional Electron Gas. *Phys. Rev. Lett.* **56**, 2104-2106 (1986).

14 Kukushkin, I. V., Smet, J. H., Scarola, V. W., Umansky, V. & von Klitzing, K. Dispersion of the Excitations of the Fractional Quantum Hall States. *Science* **324**, 1044-1047 (2009).

15 Stotz, J. A. H., Hey, R., Santos, P. V. & Ploog, K. H. Coherent spin transport through dynamic quantum dots. *Nature Mater.* **4**, 585-588 (2005).

16 Metcalfe M., Carr S. M., Muller, A., Solomon G. S. & Lawall J. Resolved sideband emission of InAs/GaAs quantum dots strained by surface acoustic waves. *Phys. Rev. Lett.* **105**, 037401 (2010).

17 Völk, S., Schülein, F. J. R., Knall, F., Reuter, D. Wieck, A. D., Truong, T. A., Kim, H., Petroff, P. M., Wixforth, A. & Krenner, H. J. Enhanced sequential carrier capture into individual quantum dots and quantum posts controlled by surface acoustic waves. *Nano Lett.* **10**, 3399-3407 (2010).

18 Krishnamurthy, S. & Santos P. V. Optical modulation in photonic band gap structures by surface acoustic waves. *J. Appl. Phys.* **96**, 1803-1810 (2004).

19 de Lima, M. M., van der Poel, M., Santos, P. V. & Hvam J. M. Phonon-Induced Polariton Superlattices. *Phys. Rev. Lett.* **97**, 045501 (2006).

20 de Lima, M. M. & Santos, P. V. Modulation of photonic structures by surface acoustic waves, *Rep. Prog. Phys.* **68**, 1639-1701 (2005).

21 Ruppert, C., Neumann, J., Kinzel, J. B., Krenner, H. J., Wixforth, A. & Betz, M. Surface acoustic wave mediated coupling of free-space radiation into surface plasmon polaritons on plain metal films. *Phys. Rev. B* **82**, 081416 (R) (2010).

22 Akahane, Y., Asano, T., Song, B.-S. & Noda, S. High-Q photonic nanocavity in a two-dimensional photonic crystal. *Nature* **425**, 944-947 (2003).

23 Berstermann, T., Brüggemann, C., Bombeck, M., Akimov, A. V., Yakovlev, D. R., Kruse, C., Hommel, D. & Bayer, M. Optical bandpass switching by modulating a microcavity using ultrafast acoustics. *Phys. Rev. B* **81**, 085316 (2010).

24 Santos, P. V. Collinear light modulation by surface acoustic waves in laterally structured semiconductors. *J. Appl. Phys.* **89**, 5060-5066 (2001).

25 Völk, S., Knall, F., Schülein, F. J. R., Truong, T. A., Kim, H., Petroff, P. M., Wixforth, A. & Krenner, H. J. Direct observation of dynamic surface acoustic wave controlled carrier injection into single quantum posts using phase-resolved optical spectroscopy. *Appl. Phys. Lett.* **98**, 023109 (2011).

26 Kress, A., Hofbauer, F., Reinelt, N., Kaniber, M., Krenner, H. J., Meyer, R., Böhm, G. & Finley, J. J. Manipulation of the spontaneous emission dynamics of quantum dots in two- dimensional photonic crystals. *Phys. Rev. B* **71**, 241304 (R) (2005).

27 Tanabe, T. Notomi, M., Kuramochi, E., Shinya, A. & Taniyama, H. Trapping and delaying photons for one nanosecond in an ultrasmall high-Q photonic-crystal nanocavity. *Nature Photon.* **1**, 49-52 (2007).

28 Liu, Z., Zhang, X., Mao, Y. Zhu, Y. Y., Yang, Z., Chan, C. T. & Sheng, P. Locally Resonant Sonic Materials. *Science* **289**, 1734-1736 (2000).

29 Mohammadi, S., Eftekhar, A. A., Hunt, W. D. & Adibi, A. High-Q micromechanical resonators in a two-dimensional phononic crystal slab. *Appl. Phys. Lett.* **94**, 051906 (2009).

30 Eichenfield, M., Chan, J., Safavi-Naeini, A. H., Vahala, K. J. & Painter, O. Modeling dispersive coupling and losses of localized optical and mechanical modes in optomechanical crystals. *Optics Express* **17**, 20078-20098 (2009).

31 Gavartin, E., Braive, R., Sagnes, I., Arcizet, O., Beveratos, A., Kippenberg, T. J. & Robert-Philip I. Optomechanical Coupling in a Two-Dimensional Photonic Crystal Defect Cavity. *Phys. Rev. Lett.* **106**, 203902 (2011).

**Supplementary information** is linked to the online version of this paper.

**Acknowledgements** This work was supported by the DFG as part of the cluster of excellence "Nanosystems Initiative Munich" (NIM) and via the Emmy-Noether-Programme (KR 3790/2-1), by the Bavaria-California Technology Center (BaCaTeC), by NSF via NIRT Grant No. 0304678 and Marie Curie EXT-CT-2006-042580. A portion of this work was done in the UCSB nanofabrication facility, part of the NSF funded NNIN network. S.M.T. acknowledges financial support from the U.S. Department of Education GAANN grant. D.A.F. acknowledges support from the Bayerische Forschungsstiftung.

**Author contributions** D.A.F. performed the experiments and 3D-FDTD simulations. D.A.F. and S.M.T. designed, fabricated and characterized the devices. H.K. and P.M.P. fabricated the MBE material. D.A.F. and H.J.K. performed the data analysis and modelling, conceived the 3D-FDTD simulations and wrote the manuscript with contributions from all other authors. H.J.K, D.B., P.M.P. and A.W. inspired and supervised the project.

**Author information** Correspondence and requests for materials should be addressed to H.J.K. (hubert.krenner@physik.uni-augsburg.de).

**Figures:**

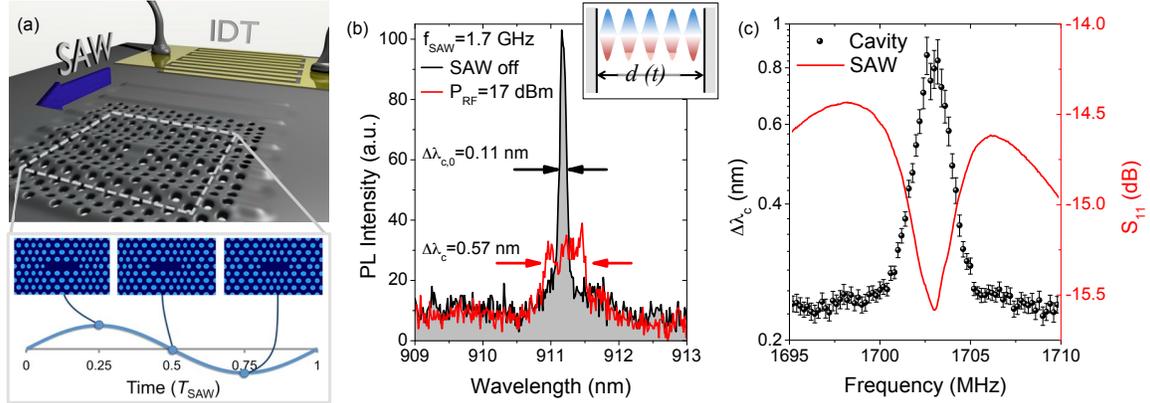

**Figure 1** – Tuning mechanism setup and experimental results. **a,** Photonic crystal nanocavity, deformed by surface acoustic wave (SAW). The SAW is generated by a RF pulse applied to an interdigital transducer (IDT). Deformation of the nanocavity at different times during one period of the SAW, showing the stretched cavity, a node at the cavity centre and the compressed cavity (not to scale). **b,** Photoluminescence from a high-$Q$ nanocavity in a time-integrated experiment. The initial Lorentzian cavity emission line (black) broadens into a line of non-trivial shape under the influence of the SAW at 1.7 GHz by more than a factor of 5 (red). Inset: Schematic of a 2.5 λ Fabry-Perot resonator. **c,** Extracted broadening of the nanocavity emission as a function of the applied RF frequency (symbols). The largest broadening is observed at the optimum operation frequency for SAW generation of the IDT (red line).

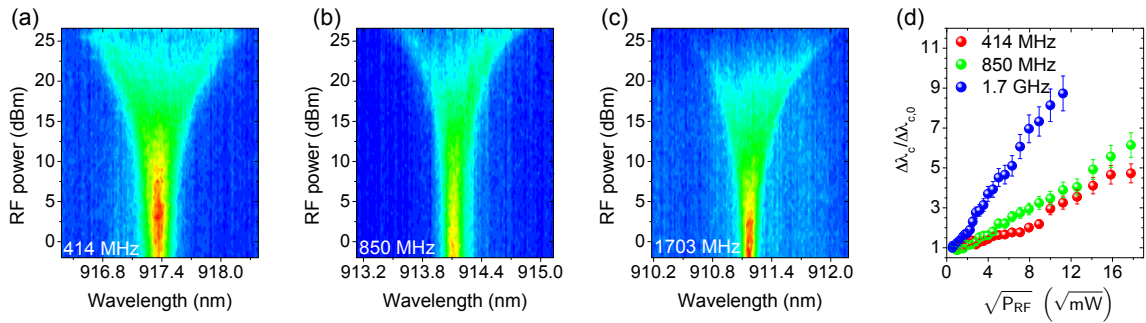

**Figure 2** – Time integrated emission spectra for three different nanocavities and SAW frequencies. **a-c,** False colour representations of cavity emission spectra as a function of the applied RF power for three different SAW frequencies. The corresponding IDT periods are 28$a$, 14$a$ and 7$a$, with $a = 0.26$ μm being the lattice constant of the photonic crystal. **d,** Spectral broadening in units of the unperturbed cavity linewidth as a function of $\sqrt{P_{RF}} \propto A_{SAW}$ showing a clear linear dependence for all three frequencies.

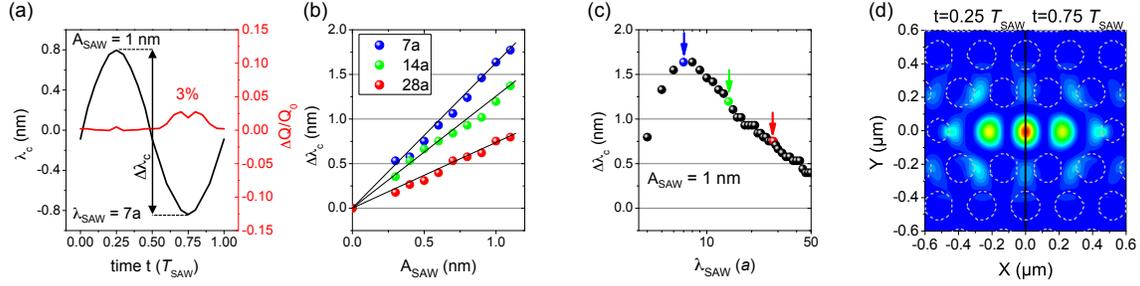

**Figure 3** – Numerical simulations. **a,** 3D-Finite Difference Time Domain simulations of the *L3* cavity show in a sinusoidal change of the mode wavelength and a minimal change of the *Q*-factor. ($\lambda_{SAW} = 7a$, $A_{SAW} = 1$ nm). **b,** Calculation of $\Delta\lambda_c$ in dependence of the SAW amplitude, showing a linear increase. **c,** The variation of the SAW wavelength at a fixed amplitude ($A_{SAW} = 1$ nm) results in an exponential increase of $\Delta\lambda_c$ with smaller wavelengths down to $\lambda_{SAW} \sim 7a$. The wavelengths investigated experimentally are marked with arrows. **d,** Comparison of the calculated spatial mode profile of the fundamental optical mode for the maximally extended and compressed nanocavity at $t = 0.25\ T_{SAW}$ and $t = 0.75\ T_{SAW}$, respectively.

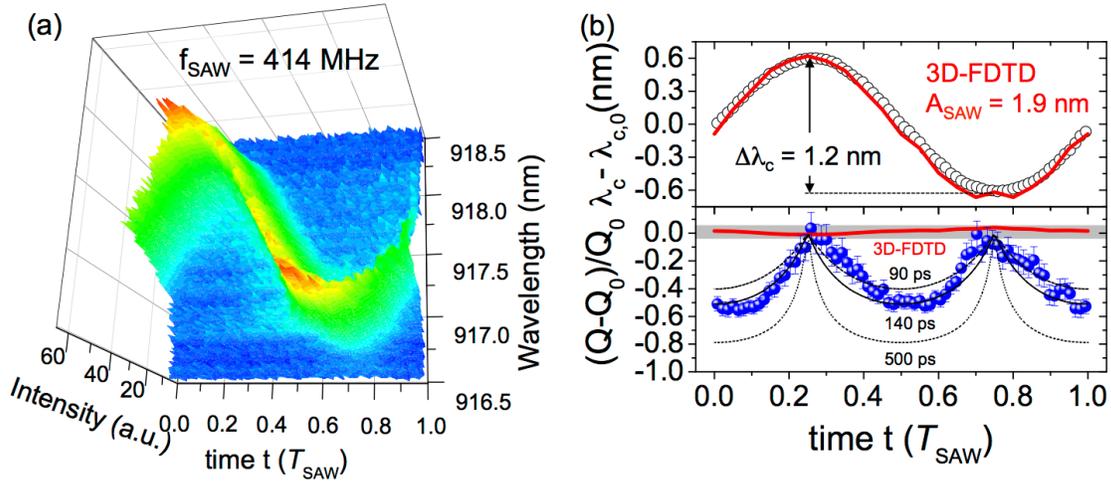

**Figure 4** – Comparison of SAW phase-resolved experimental data with 3D-FDTD simulation results. **a,** Phase-correlated stroboscopic spectroscopy measurement of the nanocavity mode emission tuned by a 414 MHz SAW. **b,** Upper panel: Extracted spectral mode position (symbols) and simulation with $\lambda_{SAW} = 28a$ and SAW amplitude of 1.9 nm. Lower panel: Estimation of the nominal *Q*-factor variation extracted from the experimental data (symbols). The black curves show the maximum Q-factor reduction resolvable for given temporal resolutions ($\tau_{laser} = 90$ ps). Agreement for the sum of the laser pulse width and Purcell-enhanced radiative lifetime ($\tau = 140$ ps) confirms minimal *Q*-factor degradation where as a longer timescale ($\tau = 500$ ps) would result to a larger contrast. The Q-factor modulation extracted from 3D-FDTD simulations is found to be <6% (red line and grey shaded bar).